\documentclass[twocolumn,aps,reprint,showpacs,raggedbottom,nobalancelastpage,amssymb,superscriptaddress,longbibliography]{revtex4-2}
\usepackage[utf8]{inputenc}
\pdfoutput=1
\usepackage{color}
\usepackage{graphicx}
\usepackage{physics}
\usepackage{braket}
\usepackage{mathtools}
\usepackage{siunitx}
\usepackage[utf8]{inputenc}
\usepackage{booktabs}
\usepackage{multirow}
\usepackage{rotating}
\newcommand{\epsd}{\epsilon_{\text{D}}^{\phantom\dag}}

\newcommand{\td}{t_{\text{d}}^{\phantom\dag}}

\newcommand{\tdl}{t_{\ell}}

\newcommand{\cdag}{c^{\dag}}
\renewcommand{\ddag}{d^{\dag}}

\newcommand{\cpdag}{c^{\phantom\dag}}

\usepackage{tikz}

\newcommand*\circled[1]{\tikz[baseline=(char.base)]{
    \node[shape=circle, draw, inner sep=0.5pt] (char) {#1};}}
\begin{document}

\title{%Coupling Kondo impurities via coherent and incoherent RKKY interactions\\
Coherent exchange-coupled nonlocal Kondo impurities \\
%A Kondo box as an entanglement bus for Kondo cat formation\\
}

\author{Lidia~Stocker}
\affiliation{Institute for Theoretical Physics, ETH Zurich, 
	8093 Zurich, Switzerland}
\author{Oded~Zilberberg}
\affiliation{Department of Physics, University of Konstanz, 78457 Konstanz, Germany}

\begin{abstract}
Quantum dots exhibit a variety of strongly correlated effects, e.g., when tuned to emulate localized magnetic impurities that form a Kondo singlet with their surrounding environment. Interestingly, in double-dots setups, the magnetic impurities couple to each other by direct Ruderman-Kittel-Kasuya-Yosida (RKKY) interaction, which wins over the Kondo physics. In this work, we investigate a double-dot device where the dots are coupled via off-resonant ballistic whispering gallery modes, dubbed electronic cavity modes. Within this cavity-double-dot system, we study, using variational matrix product state techniques, the competition between Kondo formation and the coherent RKKY-like interaction that the cavity facilitates. Specifically, we find that (i) Kondo can win and form on each dot individually, or (ii) the cavity can win and mediate between the two dots either a singlet or a novel nonlocal Kondo-like effect phase, which we call ``cat Kondo''. We systematically study the quantum phase transitions between the different many-body states. Our discoveries lay the foundation for the experimental observation of unconventional nonlocal magnetic impurities.
\end{abstract}
\maketitle

%%%%%%%% Introduction
\textit{Introduction.} The study of quantum dot systems draws continuous activity in the field of condensed matter physics, due to their potential applications in quantum information processing~\cite{loss_quantum_1998,burkard_semiconductor_2023}, as well as their tunable ability to explore strongly correlated effects~\cite{glazman_kondo_2000,coleman_introduction_2015}. A prominent example of strongly correlated physics in quantum dots is the Kondo effect~\cite{kondo_resistance_1964}. It manifests when a dot's electron acts as a spin-degenerate magnetic impurity that is screened by the surrounding environment, leading to the formation of a macroscopic dot-environment spin singlet~\cite{abrikosov_magnetic_1969, glazman_kondo_2000, ng_-site_1988, kawabata_electron_1991}. In the case of double-dot systems, an orbital-degeneracy variant known as the charge/orbital Kondo effect emerges.
This leads to the formation of a many-body state that effectively screens the degenerate charge configuration spinor~\cite{amasha_pseudo_2013}. Apart from Kondo-like effects, double-dot systems are interesting due to the Ruderman-Kittel-Kasuya-Yosida (RKKY) interaction, which mediates effective coupling between distant impurities~\cite{ruderman_indirect_1954,kasuya_theory_1956,yosida_magnetic_1957}. Combined, the two opposing effects compete: the Kondo effect tends to screen local moments, and the RKKY interaction tends to order local moments. Understanding this competition holds immense implications for comprehending correlated electron systems and has been widely investigated in a variety of systems~\cite{jayaprakash_two-impurity_1981,jones_low-temperature_1988,utsumi_rkky_2005, vavilov_transport_2005,vernek_spin-polarized_2013,bayat_entanglement_2012,pan_kondo_2016,weymann_su4_2018,parafilo_tunable_2018}.

The RKKY interaction can also serve as a knob for quantum applications~\cite{piermarocchi_optical_2002}. Here, one seeks to coherently control the spin states of the dots and implement local quantum operations, while keeping the ability to transfer quantum information between the system's building blocks. To this end, it is useful to keep the quantum dots separate in order to allow for precise control and manipulation of their individual properties~\cite{divincenzo_topics_1997}. Note, however, that in separated double dot systems, the central separating lead harbors partially suppressed RKKY interaction that coexists with a superexchange interaction~\cite{ong_generalized_2011}. This complicates its harnessing as a coherent entanglement bus. As such, there is a variety of alternative proposals for coupling distant quantum dots, e.g., via the edge modes of the quantum Hall effect~\cite{yang_quantum-hall_2002,scarola_possible_2002,elman_long_2017}, using superconducting cavities~\cite{petersson_circuit_2012,toida_vacuum_2013,liu_photon_2014,deng_charge_2015,stockklauser_strong_2017,gu_probing_2023}, or replacing the central lead with a large, yet interacting dot~\cite{craig_tunable_2004}. 

Interestingly, coherent coupling between distant dots was experimentally achieved using a large open dot that has a structured density of states~\cite{nicoli_cavity-mediated_2018}. The structure harbors ballistic standing waves that are embedded in the larger expanse of states in the system~\cite{rossler_transport_2015,ferguson_long-range_2017,dias_da_silva_conductance_2017,gold_imaging_2021}. This is the mesoscopic equivalent of quantum corrals~\cite{katine_point_1997,hersch_diffractive_1999}, which we dub \textit{electronic cavity states}. The first experimental realization of an electronic cavity coupled to a single dot showed a competition between strongly correlated effects, namely, ``molecular'' dot-cavity singlet formation competed with the Kondo effect~\cite{rossler_transport_2015,ferguson_long-range_2017,stocker_entanglement-based_2022}. These results laid the foundation for applying all-electronic dot-cavity devices not only for quantum information processing applications but as an interesting platform for the fundamental study of strongly correlated physics.

In this work, we present a comprehensive study of the cavity-mediated strongly-correlated states in a separated double-dot device. We observe the emergence of both RKKY-like interaction between distant dots and nonlocal spin-Kondo formation between the double-dot and the environment. Our study involves exact numerical techniques, which reveal the order parameters (fingerprints) of the different states, as well as the crossover between them. These fingerprints can be observed using tomography of the double dot system~\cite{stocker_entanglement-based_2022}. When possible, our results are supported by analytical derivation of the order parameters. First, we provide an exact solution to the interplay among Kondo, RKKY, and ferromagnetic interactions in a double-dot system coupled to a central continuous lead~\cite{ong_generalized_2011}. Next, we delve into the coherent physics that arises when the system is coupled to a detuned cavity. Here, we capture variants of the standard Kondo and RKKY effect, alongside predicting a novel exchange-based nonlocal Kondo effect. Our results provide a comprehensive map of the many-body effects arising in separated double dots and inspire their experimental realization.

\begin{figure}
\centering\includegraphics[width=8.6cm]{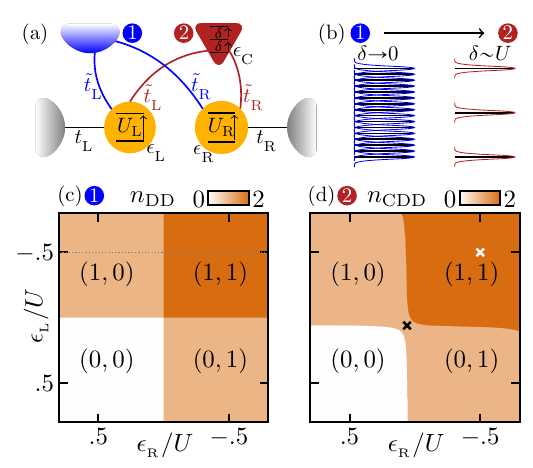}
	\caption{\textit{System and charge stability diagram.} (a) Double-dot system [cf.~Eq.~\eqref{eq: effetive model double dot}] composed of two spinful single-level quantum dots, $\ell{\in}\text{L,R}$ (yellow circle) with on-site energy $\epsilon_{\ell}$ and  interaction $U_\ell$. They are tunnel-coupled (black lines) to their respective leads (grey semicircles) with tunnelling amplitudes $t_\ell$. The two dots are additionally tunnel-coupled (blue/red lines) to a central common lead with amplitudes $\tilde{t}_\ell^{\phantom\dag}$. (b) The central box has energy spacing $\delta$ with two limiting cases: (1, blue) a metallic lead when $\delta{\to}0$, and (2, red) an electronic cavity when $\delta{\sim}U$. In reality~\cite{nicoli_cavity-mediated_2018}, the electronic cavity is open, as depicted by broadened levels. (c) Charge stability diagram of the double-dot system (1). (d) Charge stability diagram of the cavity-double-dot system (2), where the cavity is truncated to a single-energy level $\epsilon_{c}{=}0.75U$ ($\delta{\to}\infty$) for $\tilde{t}_{\ell}{=}0.1U$. As the cavity is detuned from the Fermi level, its occupation is $n_{_\text{C}}{\approx}0$. The population of the two dots is marked by $(n_{\text{L}},n_{\text{R}})$. The grey dotted line and white and black $\cross$ mark the regions discussed later in the manuscript, see Figs.~\ref{fig: figure 2}-\ref{fig: figure 3}.}
	\label{fig:model}
\end{figure}

%%%%%%%%%%%%%%%%% Model
\textit{Model.}
Our double-dot system is composed of two Anderson impurity models~\cite{anderson_localized_1961}, each coupled to its own lead (environment), as well as to a central common lead, see Fig.~\ref{fig:model}(a). Its effective model, derived in Ref.~[\onlinecite{ferguson_long-range_2017}], reads
\begin{align}
\resizebox{0.9\hsize}{!}{$\displaystyle{H  = \sum_{\ell}\left(H_{\ell} + H_{\text{lead}}^\ell\right) + H_{\text{C}}^{\delta} + \sum_\ell \left(H_{\text{tun}}^{\text{lead}_\ell{-}\ell} + H_{\text{tun}}^{\text{C}{-}\ell}\right) \ ,}$}
\label{eq: effetive model double dot}
\end{align}
where $\ell{\in}\left\{\text{L},\text{R}\right\}$ denotes the left and right dot, respectively. Each dot Hamiltonian
$H_{\ell}{=}\sum_\sigma^{\phantom \dag}\epsilon_{\ell}^{\phantom\dag}n_{\ell\sigma}^{\phantom\dag}{+}U_\ell^{\phantom\dag} n_{\ell\uparrow}^{\phantom\dag}n_{\ell\downarrow}^{\phantom\dag}$
describes an impurity with a spin-degenerate electron level at energy $\epsilon_{\ell}$, and electron-electron charging energy $U_\ell^{\phantom\dag}$. Here, $n_{\ell\sigma}^{\phantom\dag}$ denotes the dot level's occupation number with spin $\sigma{\in}\{\uparrow,\downarrow\}$. The left and right leads are noninteracting continuous reservoirs
$H_{\text{lead}}^{\ell}{=}\sum_{k\sigma}^{\phantom\dag}\epsilon_{k\ell}^{\phantom\dag}\cdag_{k\ell\sigma}\cpdag_{k\ell\sigma}$, where we denote $c_{k\ell\sigma}^{\phantom\dag}$ ($\cdag_{k\ell\sigma})$ as the fermionic annihilation (creation) of an electron with momentum $k$ and spin $\sigma$ in the $\ell^{\rm th}$ lead. Each dot is coupled to its own lead via $H_{\text{tun}}^{\text{lead}_\ell{-}\ell}{=}\sum_{\ell  k\sigma}\tdl\ddag_\sigma \cpdag_{k\ell\sigma}{+}\text{H.c.}$ with energy-independent tunnelling amplitudes $t_{\ell}$. We consider the central region as a set of noninteracting and equally spaced energy levels $H_{C}^{\delta}{=}\sum_{j\sigma}\left(\epsilon_{_\text{C}}{+}j\delta\right)c^{\dag}_{j\sigma}c_{j\sigma}$, where the fermionic operators $c_{j\sigma}^{\phantom\dag}$ ($\cdag_{j\sigma})$ are defined as those of the left and right leads. The central region is tunnel-coupled to both dots $H_{\text{tun}}^{C{-}\ell}{=}\sum_{\ell  j\sigma}\tilde{t}_{\ell}\ddag_{\ell\sigma} \cpdag_{j\sigma}{+}\text{H.c.}$ with energy-independent tunnelling amplitudes $\tilde{t}_{\ell}$. 
In the limit of vanishing level spacing $\delta{\to}0$, denoted as system (1), the central region corresponds to a lead, see Fig.~\ref{fig:model}(b). In the situation where $\delta{\sim}U$, denoted as system (2), the levels of the central region are discrete and correspond to a multimode electronic cavity, used in the Kondo-box problem~\cite{thimm_kondo_1999}.
In the following, for clarity, we consider identically tuned dots $U_\ell{\equiv}U, t_{\ell}{\equiv}t_{\text{D}}, \tilde{t}_{\ell}{\equiv}\tilde{t}_{\text{D}}$.

We begin by analysing the charge stability diagram of the double-dot system in the two limiting situations, namely (1) the separated double-dot system, and (2) the cavity-double-dot system, see Figs.~\ref{fig:model}(c) and (d). We assume that coupling to the leads is vanishing $t_\ell{\approx}\tilde{t}_\ell{\approx}0$, and exactly diagonalize the remaining ``closed'' impurity system. In case (1), the diagram of the closed two-dot system $H_{\text{DD}}{=}\sum_{\ell}H_{\ell}$, with $\langle n_{\text{DD}}\rangle{=}\langle n_{\text{L}}{+}n_{\text{R}}\rangle$, exhibits standard Coulomb blockade on each of the dots individually, separated by ``resonance lines'' where the particle number on the double-dot is ill-defined. In case (2), the closed system is composed of the two dots and the discrete levels defining the cavity. Here and in the following, we truncate and consider a cavity with a single energy level. Coupling the cavity to the dots, $\tilde{t}_{_\text{D}}{\neq}0$, we obtain an ``artificial molecule'' with Hamiltonian $H_{\text{CDD}}{=}\sum_{\ell}\left(H_{\ell}{+}H_{\text{tun}}^{C{-}\ell}\right){+}H_{\text{C}}^{\delta}$, which for $\left|\tilde{t}_\ell\right|{>}0$ creates avoided crossings as the levels with same total number of electrons $\langle n_{\text{CDD}}\rangle{=}\langle n_{\text{C}}{+}n_{\text{L}}{+}n_{\text{R}}\rangle$ hybridize via the cavity.

\begin{figure*}
\centering\includegraphics[width=17.4cm]{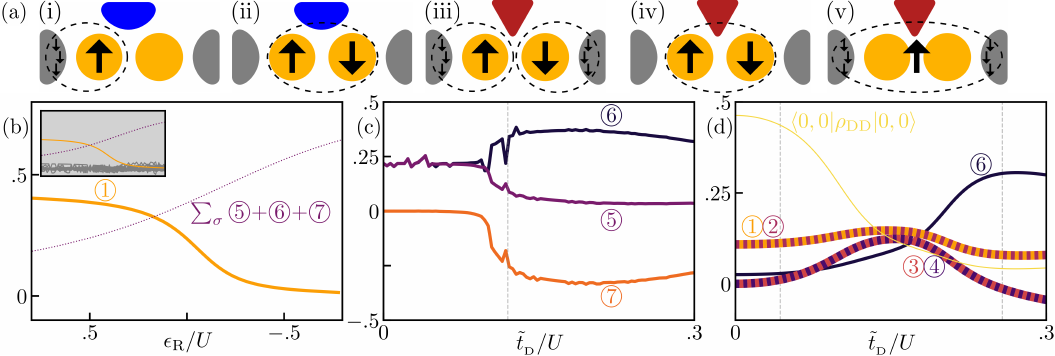}
	\caption{\textit{Crossovers between ground states.} (a) Competing hybridization  mechanisms (dashed circle): (i) Standard spin Kondo between the left dot and its lead, (ii) dot-dot hybridization via the central lead, (iii) two separate dot-lead spin Kondo, (iv) cavity-mediated RKKY dot-dot singlet, and (v) Kondo cat. (b)-(d) NRG-MPS tomography results of  the reduced double-dot impurity matrix elements $\bra{s,s'}\rho_\text{DD}\ket{s'',s'''}$. Out of the 256 matrix elements, we plot those with a contribution $\left|\bra{s,s'}\rho_\text{DD}\ket{s'',s'''}\right|{>}0.1$ in the interval considered in each respective plot. The circular numbered markers correspond to the elements in Table~\ref{table 1}. The line width represents the degeneracy of the configuration; one-fold: thin, two-fold: standard, four-fold: thick.  (b) System (1) with $\epsilon_{_\text{L}}{=}{-}U/2$, $\tilde{t}_\text{D}{=}0.1$. (inset) all the 256 matrix elements are plotted. Note that we sum over the (anti)ferromagnetic contributions (dotted line), cf. discussion in the main text. (c) System (2) with $\epsilon_{_\text{L}}{=}\epsilon_{_\text{R}}{=}{-}U/2$ [cf.~white $\cross$ in Fig.~\ref{fig:model}(d)]. Grey line marks the boundary between the left region (where the Kondo gap dominates the RKKY gap) and the right region (where the RKKY gap dominates the Kondo gap). (d) System (2) with $\epsilon_{_\text{L}}{=}\epsilon_{_\text{R}}{=}0.005U$ [cf. black $\cross$ in Fig.~\ref{fig:model}(d)]. Grey lines mark the regions with a different total number of electrons for the closed cavity-double-dot system, $n_\text{DD}{\approx}2,1,0$. We set the dot-leads coupling $t_\text{D}{=}0.1U$, the NRG chain length of each lead $N{=}80$, the MPS bond dimension $D{=}500$, and the leads are assumed to have a constant density of states $d_0{=}1/(2U)$.
 }
	\label{fig: figure 2}
\end{figure*}

\textit{Methodology.}
To analyze the many-body physics in our system, we calculate the ``ground state'' of the open system (impurity plus leads) using a numerical NRG-MPS method~\cite{wilson_renormalization_1975,krishna-murthy_renormalization-group_1980,krishna-murthy_renormalization-group_1980-1,bulla_numerical_2008,saberi_matrix-product-state_2008,weichselbaum_variational_2009,stocker_entanglement-based_2022}. We consider the system at equilibrium (zero bias voltage $\mu_\text{L}{=}\mu_\text{R}{=}\mu_\text{C}$).
We, then, trace out the leads and extract the elements of the (double-dot) impurity's reduced $16{\times}16$ density matrix, $\rho_\text{DD}$. These 256 density matrix elements act as witnesses (order parameters) for the type of strongly correlated states that form between the dots and their leads, and can be probed using tomography methods~\cite{stocker_entanglement-based_2022}. In analytically treatable cases, we harness an analytical Schrieffer-Wolff transformation (SWT)~\cite{schrieffer_relation_1966, bruus_many-body_2004} to predict which values of $\rho_\text{DD}$ are markers of which many-body state. In Table~\ref{table 1}, we summarize the (nonvanishing) values of the density matrix elements $\left|\bra{s,s'}\rho_\text{DD}\ket{s'',s'''}\right|$ corresponding the variety of effects (i)-(v) identified in this work~\cite{supmat}, where $\ket{s,s'}{=}\ket{s}_\text{L}{\otimes}\ket{s'}_\text{R}$ and $s,s'{\in}\left\{0,\uparrow,\downarrow,\uparrow\downarrow\right\}$ denote the spin configuration of the left and right dot. The (i)-(v) effects are depicted in Fig.~\ref{fig: figure 2}(a). Note that we apply the NRG-MPS tomography method to both cases (1) and (2). In the former, we introduce an additional lead to the environment by changing the NRG-MPS decomposition.

\begin{table}
\begin{ruledtabular}
\begin{tabular}{cccccc}
  & (i)&  (ii)&  (iii)&(iv) & (v)  \\
 \hline
  \circled{1}  $\bra{0,\sigma}\rho_\text{DD}\ket{0,\sigma}$&  .5&  0&  0&  0& .25 \\
   \circled{2} $\bra{\sigma,0}\rho_\text{DD}\ket{\sigma,0}$ &  0&  0&  0&  0& .25 \\
     \circled{3}   $\bra{\sigma,0}\rho_\text{DD}\ket{0,\sigma}$&  0&  0&  0&  0& .25 \\
  \circled{4} $\bra{0,\sigma}\rho_\text{DD}\ket{\sigma,0}$&  0&  0&  0&  0& .25 \\
  \circled{5} $\bra{\sigma,\sigma}\rho_\text{DD}\ket{\sigma,\sigma}$ &0  & \multirow{3}{*}{\hspace{-0.05em}$\left.\begin{array}{l}
                \\
                \\
                \\
                \end{array}\right\rbrace\sum{=}1$ } &  .25& 0& 0\\
 \circled{6}  $\bra{\sigma,\bar{\sigma}}\rho_\text{DD}\ket{\sigma,\bar{\sigma}}$ &0  &  & .25 & .5&0 \\
   \circled{7}  $\bra{\sigma,\bar{\sigma}}\rho_\text{DD}\ket{\bar{\sigma},\sigma}$ &0  &  &  0& -.5& 0\\
  \end{tabular}
\caption{The tomography order parameters corresponding to the (i)-(v) effects illustrated in Fig.~\ref{fig: figure 2}(a), as calculated by exact diagonalization and SWT~\cite{supmat}. We omit the other 249 density matrix elements, as they are $=0$ in all of the cases considered.}\label{table 1}
\end{ruledtabular}
\end{table}
\textit{Double-dot system; case (1).} As an example for our procedure, we first consider the case where $\epsilon_{_\text{L}}{=}{-}U/2$, and tune the level of the right dot  $\epsilon_{_{\text{R}}}$, see Fig.~\ref{fig: figure 2}(b). Here, the left dot is singly occupied, while the right moves from being empty to singly occupied, as $\epsilon_{_{\text{R}}}$ decreases, cf.~Fig.~\ref{fig:model}(c). For the double-dot impurity, our NRG-MPS tomography procedure produces 256 values, see inset of Fig.~\ref{fig: figure 2}(b). The significant order parameters are few, and we filter out small values $\left|\bra{s,s'}\rho_\text{DD}\ket{s'',s'''}\right|{<}0.1$. As the right dot becomes more occupied, the order parameter $\bra{0,\sigma}\rho_\text{DD}\ket{0,\sigma}$ decreases, whereas another order parameter $\sum_\sigma\bra{\sigma,\sigma}\rho_\text{DD}\ket{\sigma,\sigma}{+}\bra{\sigma,\bar{\sigma}}\rho_\text{DD}\ket{\sigma,\bar{\sigma}}{+}\bra{\sigma,\bar{\sigma}}\rho_\text{DD}\ket{\bar{\sigma},\sigma}$ grows to become dominant. The values of the former, cf. Table~\ref{table 1}, are a fingerprint for a Kondo singlet formation between the left dot and its lead, as predicted from standard SWT~\cite{glazman_kondo_2000,supmat}, cf.~Fig.~\ref{fig: figure 2}(a,i). The values of the latter identify dot-dot hybridization mediated by the central-lead, where ferromagnetic superexchange coexists with antiferromagnetic RKKY-like interactions~\cite{ong_generalized_2011, supmat}, cf.~Fig.~\ref{fig: figure 2}(a,ii). Note that we depict the sum over the contributions of the (anti)ferromagnetic terms due to the degenerate ground state in the $\epsilon_{_\text{L}}{<}0$ case~\cite{supmat}. 
Such a transition was observed experimentally~\cite{craig_tunable_2004} and discussed analytically~\cite{ong_generalized_2011} in different double-dots setups. Here, we move beyond perturbative approaches and capture these effects (including the crossover between them) using a numerically exact method on the full (open) system. This is the first key result of our work.

\textit{Cavity-mediated RKKY; case (2).} We consider now the cavity-double-dot system in the $\epsilon_{_\text{L}}{=}\epsilon_{_\text{R}}{=}{-}U/2$ regime, where both left and right dot are singly occupied. In Fig.~\ref{fig: figure 2}(c), we tune the dots-cavity coupling $\tilde{t}_\text{D}$ and plot the filtered tomography values. As $\tilde{t}_\text{D}$ increases, the term $\bra{\sigma,\bar{\sigma}}\rho_\text{DD}\ket{\sigma,\bar{\sigma}}$ increases, while the term $\bra{\sigma,\sigma}\rho_\text{DD}\ket{\sigma,\sigma}$ decreases towards zero and the term $\bra{\sigma,\bar{\sigma}}\rho_\text{DD}\ket{\bar{\sigma},\sigma}$ appears. The values of the former two order parameters are the fingerprints of Kondo singlets forming between each dot and their respective lead independently, cf. Table~\ref{table 1}, and Fig.~\ref{fig: figure 2}(a,iii). As they gap out and the third term appears, we observe the fingerprint of a cavity-mediated singlet that forms on the two dots~\cite{supmat}, cf. Table~\ref{table 1}, and Fig.~\ref{fig: figure 2}(a,iv). Crucially, the singlet can be attributed to a ``coherent RKKY interaction'', unlike the standard RKKY mechanism that involves a continuum. Such coherent hybridization  engenders a so-called exchange blockade~\cite{nicoli_cavity-mediated_2018}, further distinguishing it from conventional RKKY behavior. 

The formation of two separated Kondo effects opens a gap that is twice that of the Kondo gap for a single dot system~\cite{supmat}
\begin{equation}
\label{eq: Kondo temperature quantum dot with one lead}
\Delta_\text{K}= \sqrt{2\pi|t_\text{D}|^2Ud_0}\exp\Biggl(\dfrac{\epsd(\epsd+U)}{2|t_\text{D}|^2Ud_0}\Biggr) \ ,
\end{equation}
where $d_0$ is the leads' density of states.
Similarly, the dot-dot singlet formation opens a gap~\cite{supmat}
\begin{equation}
    \Delta_\text{RKKY} \approx 48\tilde{t}_{_\text{D}}^4{/}\left(U{+}2\epsilon_{_\text{C}}\right)^3 \ .
    \end{equation}
 The vertical grey line in Fig.~\ref{fig: figure 2}(c) marks the critical $\tilde{t}_{\text{D}}$ value above which the RKKY dominates the Kondo gap $\Delta_\text{RKKY}{>}\Delta_\text{K}$ and therefore a dot-dot singlet formation is expected. This value is in good agreement with the crossover observed with our NRG-MPS tomography results discussed above. A signature of such a dot-dot singlet formation was experimentally detected~\cite{nicoli_cavity-mediated_2018}, with a theoretical description that was limited to the closed system. Here, we predict that the singlet can form in the realistic many-body setting and win against competing hybridization channels with the leads. It would be interesting to experimentally tune the cavity level $\epsilon_c$ in this detuned regime, and observe the appearance of Kondo singlets. This is the second key result of our work; we numerically resolve the coherent long-range coupling between distant dots in the complex open cavity-double-dots system.

\textit{Kondo cat.} 
As shown in Fig.~\ref{fig:model}(d), the coupling to the cavity opens a gap in the $\epsilon_{_\text{L}}{\approx}\epsilon_{_\text{R}}{\gtrsim}0$ region and the total occupation of the double dot system is $n_\text{DD}{\approx}1$ in the exchange blockade regime. We now set the dots' energy levels to $\epsilon_{_\text{L}}{=}\epsilon_{_\text{R}}{=}0.005U$, and tune the dots-cavity coupling strength $\tilde{t}_\text{D}$, see Fig.~\ref{fig: figure 2}(d). The empty-dots term $\bra{0,0}\rho_\text{DD}\ket{0,0}$ decreases towards zero for increasing dots-cavity coupling as the exchange-gap opens and the dots become more occupied. Conversely, the term $\bra{\sigma,\bar{\sigma}}\rho_\text{DD}\ket{\sigma,\bar{\sigma}}$ increases as both dots become singly occupied. In the midst of the parameter scan, terms with total double-dot occupation close to one $\bra{0,\sigma}\rho_\text{DD}\ket{0,\sigma}$, $\bra{\sigma,0}\rho_\text{DD}\ket{\sigma,0}$, $\bra{\sigma,0}\rho_\text{DD}\ket{0,\sigma}$, and $\bra{0,\sigma}\rho_\text{DD}\ket{\sigma,0}$ are dominant. The observed value of these order parameters are the fingerprint of an interesting nonlocal Kondo configuration, see Table~\ref{table 1} and Figs.~\ref{fig: figure 2}(a,v). This nonlocal Kondo singlet formation is established because of cavity-mediated exchange interaction over a large spatial extent. Indeed, the cavity mediates an orbital hybridization (superposition) between the dots with a sufficiently large gap to allow for the formation of this nonlocal Kondo singlet with both leads. This can be understood as a superposition of Kondo singlets forming on both dots, motivating the name Kondo cat.

To better understand the  Kondo cat, we consider the exchange blockade, appearing in the $\epsilon_\ell{\approx}0$ regime (we henceforth consider $\epsilon_{_{\text{L}}}{=}\epsilon_{_{\text{R}}}$), opens a gap~\cite{supmat}
\begin{equation}
\Delta_{\text{EXC}} = 2\tilde{t}_{_\text{D}}^2/(\epsilon_{_\text{C}}-\epsilon_\ell) \ ,  
\end{equation}
leading to $n_{\text{D}}{\approx}1$ for $\epsilon_\ell - \Delta_{\text{EXC}} < 0$, when the energy is lower than the energy of the empty state [see left vertical grey line in Fig.~\ref{fig: figure 2}(d) above which the relation holds]. From the other side, we have $\epsilon_{_\text{D}} - \Delta_{\text{EXC}} <  \min(2\epsilon_{_\text{D}} - \Delta_\text{K} , 2\epsilon_{_\text{D}} - \Delta_\text{RKKY})$, i.e., when the energy is lower than the energy of the doubly-occupied state [see right vertical grey line in Fig.~\ref{fig: figure 2}(d)  below which the second relation holds]. For the latter relation, we take into account that in the doubly-occupied region standard Kondo effects compete with the RKKY interaction. Additionally, we find that the ``ground state'' of the system in the $n_{\text{D}}{\approx}1$ configuration is spin-degenerate~\cite{supmat} and, therefore, can reduce its energy by forming a spin-Kondo singlet with the surrounding leads. Our NRG-MPS tomography results capture such a nonlocal Kondo formation in the midst of the $n_{\text{D}}{\approx}1$ regime. With contemporary control over both the cavity and the dots levels, the Kondo cat is within experimental reach with implications, and can be served as a meter for the spin coherence length scales in the system~\cite{nicoli_cavity-mediated_2018}. 
This is the third key result of our work; we predict a novel Kondo effect, where a nonlocal magnetic impurity is screened by the environment. %As the Kondo screening is macroscopic and the double-dot system in orbital superposition, we dub this configuration a \textit{Kondo cat}.
The distinct Kondo cat formation fundamentally sets our system apart from standard double-dot systems~\cite{craig_tunable_2004}.

\begin{figure}
\centering\includegraphics[width=8.7cm]{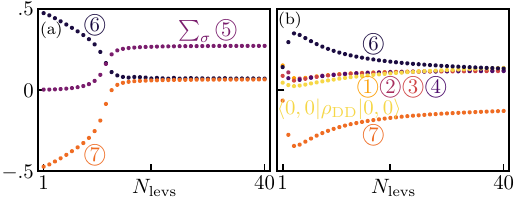}
	\caption{\textit{Crossovers between the (2) and (1) limits.} NRG-MPS tomography results of the cavity double-dot system as a function of the number of cavity energy levels $N_\text{levs}$, $\delta{=}0.25U/N_\text{levs}$. (a) System (2) with $\epsilon_{_\text{L}}{=}\epsilon_{_\text{R}}{=}{-}U/2$, $\tilde{t}_{_\text{D}}{=}0.15U$. (b) System (2) with $\epsilon_{_\text{L}}{=}\epsilon_{_\text{R}}{=}0.005U$, $\tilde{t}_{_{\text{D}}}{=}0.15U$. We use NRG chain length  $N{=}40$ and MPS bond dimension $D{=}300$. (Other parameters and markers are as in Fig.~\ref{fig: figure 2}).
 }
	\label{fig: figure 3}
\end{figure}

\textit{Dependence on the central region's level spacing.}
We have observed two distinct effects associated with the coherent dot-dot coupling as mediated by the cavity [case (2)], namely, the cavity-mediated RKKY and Kondo cat regimes. We now turn to examine the crossover between the limiting scenarios (2) and (1) in these regimes, see Fig.~\ref{fig: figure 3}. We set equidistant cavity levels $N_\text{levs}{=}1,2,\ldots$, where the spacing is determined by $\delta{=}0.25U/N_\text{levs}$~\footnote{This choice is motivated the fact that we have $\epsilon_{_\text{C}}{=}0.75U$ and aim to keep the cavity levels within the energy bandwidth of the NRG lead $\mathcal{D}{=}\left[-U,U\right]$}. In Fig.~\ref{fig: figure 3}(a), we consider the regime where, in the $N_\text{levs}{=}1$ limit, the cavity-mediated RKKY effect (iv) is observed. For increasing number of cavity levels, we observe that the antiferromagnetic order parameters go towards zero. Concurrently, the ferromagnetic terms become more dominant. Therefore, we observe how the RKKY effect is suppressed by ferromagnetic superexchange, cf. Table~\ref{table 1}. Differently from the $\delta{\approx}0$ case, the antiferromagnetic order parameters are negligible. In Fig.~\ref{fig: figure 3}(b), we consider the regime where, in the $N_\text{levs}{=}1$ limit, the Kondo-cat (v) is observed. For increasing number of cavity levels, we observe that the characteristic values of the Kondo-cat-like parameters rapidly decrease. Concurrently, the values characteristic  to the RKKY-like interaction increase. Therefore, we observe how the Kondo cat effect is suppressed by the RKKY interaction, cf. Table~\ref{table 1}. Notably, the observed crossovers occurs at $N_\text{levs}{<}10$ for both regimes. This number of levels is significantly lower than in the continuous limit $N_\text{levs}{\to}\infty$ of system (1).

\textit{Conclusion and outlook.} We find a rich variety of many-body states  within a double-dot system, with a particular focus on the competition of Kondo with RKKY-like effects and predict a novel nonlocal Kondo impurity. We harness and demonstrate the potential of tomography analysis~\cite{stocker_entanglement-based_2022} in understanding and distinguishing between the different strongly correlated states. To accomplish this, we apply and expand the NRG-MPS methodology to encompass complex multi-impurities multi-reservoir setups. Throughout the work, we employ typical values for the $\epsilon_{_\text{C}}, U, T_K$ parameters that enable the experimental exploration of the RKKY regime and detection of the novel Kondo cat regimes in the cavity-double-dot setup~\cite{rossler_transport_2015,nicoli_cavity-mediated_2018}. Future work will focus on finding transport observables sensitive to the different many-body ground states~\cite{stocker_2023}. Our findings motivate the potential application of Kondo-box-like-double-dot systems as quantum simulators and quantum information processors, potentially extending to state-of-the-art gate-defined quantum dots in silicon~\cite{weber_spin_2014} or bilayer graphene~\cite{kurzmann_kondo_2021, banszerus_tunable_2021} devices. Furthermore, the proposed tomography procedure is not limited to mesoscopic impurity setups but can also be analogously applied for exploring strongly correlations in optomechanical~\cite{aspelmeyer_cavity_2014} or cold atoms~\cite{bloch_quantum_2012} systems.

\begin{acknowledgements} We thank G. Blatter, T. Ihn, K. Ensslin, and in particular M. Ferguson for illuminating discussions, and acknowledge financial support from the Swiss National Science Foundation (SNSF) through project 190078, and from the Deutsche Forschungsgemeinschaft (DFG) - project number 449653034. Our numerical implementations are based on the ITensors \textsc{Julia} library~\cite{fishman_itensor_2022}.\
\end{acknowledgements} 

\appendix
\
\newpage
\cleardoublepage
\setcounter{figure}{0}

{\onecolumngrid
\begin{center}
	\textbf{\normalsize Supplemental Material: Coherent exchange-coupled nonlocal Kondo impurities}\\
	\vspace{3mm}
	\vspace{4mm}
	
	%{ Lidia~Stocker$^{1}$  and Oded Zilberberg$^{2}$}\\
	%\vspace{1mm}
	%\textit{\small $^{1}$Institute for Theoretical Physics, ETH Z\"urich, 8093 Z\"urich, Switzerland\\
	%}
	%\textit{\small $^{2}$Department of Physics, University of Konstanz, 78457 Konstanz, Germany}
	
	%\vspace{5mm}
\end{center}}
%%%%%%%%%% Merge with supplemental materials %%%%%%%%%%
%%%%%%%%%% Prefix a "S" to all equations, figures, tables and reset the counter %%%%%%%%%%
\twocolumngrid

\setcounter{equation}{0}
\setcounter{section}{0}
\setcounter{figure}{0}
\setcounter{table}{0}
\setcounter{page}{1}
\makeatletter
\renewcommand{\bibnumfmt}[1]{[#1]}
\renewcommand{\citenumfont}[1]{#1}
%%%%%%%%%% Prefix a "S" to all equations, figures, tables and reset the counter %%%%%%%%%%

\setcounter{enumi}{1}
\renewcommand{\theequation}{S\arabic{section}.\arabic{equation}}
\renewcommand{\thesection}{S\arabic{section}}
\renewcommand{\thetable}{S\arabic{section}.\arabic{table}}
\renewcommand{\thefigure}{S\arabic{section}.\arabic{figure}}

\section{Schrieffer-Wolff transformation }
In the following, we sketch the standard steps for applying the Schrieffer-Wolff transformation (SWT) in quantum dot systems. For a detailed derivation, see Refs.~\cite{schrieffer_relation_1966, bruus_many-body_2004}. We consider the general effective model of an impurity coupled to leads
\begin{equation}
    H = H_{\text{imp}} + H_{\text{leads}} + H_{\text{tun}}^{\text{imp-leads}} \ ,
\end{equation}
with Hamiltonian terms $H_{\text{imp}}$, $H_{\text{leads}}$, and $H_{\text{tun}}^{\text{imp-leads}}$ describing the impurity, leads, and their coupling, respectively.

The SWT involves a canonical unitary transformation $S$ that cancels the tunnelling impurity-leads terms $H_{\text{tun}}^{\text{imp-leads}}$ to lowest-order. Any canonical transformation on $H$ acts as
\begin{equation}
H_S = e^{iS}He^{-iS} =H + i[S, H] -\dfrac{1}{2}\bigl[S,[S,H]\bigr] + \ldots \ . \label{eq: Hamiltonian transformation S}
\end{equation}
Choosing $S$ to be linear in $H_{\text{tun}}^\text{imp-leads}$, we have $[S,H_{\text{tun}}^\text{imp-leads}] {=} 0$. Hence, the $S$ does not produce any linear contribution in $H_{\text{tun}}^\text{imp-leads}$. Furthermore, it is chosen such that its other commutation terms cancel any direct impurity-leads hopping terms arising in $H$. Thus, the rotated model involves only virtual transitions between impurity and leads, e.g., the second order terms in the expansion, denoted as $H_S^{(2)}$, take the form
\begin{equation}
\label{eq: SW result second order term}
H_{S}^{(2)} = J\vec{S}_\text{imp}\cdot\vec{S}_{\text{leads}} + H_{\text{scat}}^{\text{lead-lead}} \ ,
\end{equation}
where ${S}_\text{imp}, {S}_{\text{leads}}$ are average spin operators corresponding to the impurity and leads, respectively [cf.~example below; Sec.~\ref{sec:singleDot}]. The first term describes a spin scattering off of the impurity, while $H_{\text{scat}}^{\text{lead-lead}}$ are potential scattering terms between the leads. The first term corresponds to the Kondo Hamiltonian~\cite{kondo_resistance_1964}.

\subsection{SWT for the single impurity Anderson model}
\label{sec:singleDot}
As an example, we show the result of applying the SWT for the single impurity Anderson model and obtain the effective Kondo Hamiltonian [cf.~Eq.~\eqref{eq: SW result second order term}]. The model reads
\begin{align}
H  = & \sum_\sigma^{\phantom \dag}\epsilon_{_\text{D}}^{\phantom\dag}n_{\text{D}\sigma}^{\phantom\dag}{+}Un_{\text{D}\uparrow}^{\phantom\dag}n_{\text{D}\downarrow}^{\phantom\dag} + \sum_{k\sigma}^{\phantom\dag}\epsilon_{k}^{\phantom\dag}\cdag_{k\sigma}\cpdag_{k\sigma} \nonumber\\
 & + \sum_{k\sigma}\td\ddag_\sigma \cpdag_{k\sigma}{+}\text{H.c.} \ ,
\end{align}
where the creation, annihilation and occupation operators are defined as in the main text. We use $\ell{=}\text{D}$, as we consider a single dot coupled to its own lead. The second-order correction term, cf.~Eq.~\eqref{eq: SW result second order term}, for the Anderson impurity Hamiltonian results in~\cite{bruus_many-body_2004}
\begin{equation}
    H_{S}^{(2)} = \sum_{kk'}J\vec{S}_\text{D}\cdot\vec{S}_{k,k'} + \sum_{kk'\sigma}Wc^\dag_{k\sigma}c^{\phantom\dag}_{k\sigma} \ ,
    \label{eq:KondoSingle}
\end{equation}
where 
\begin{subequations}
\begin{align}
    \vec{S}_{\text{D}} & =\frac{1}{2}\sum_{\sigma\sigma'}d_{\sigma}^\dag\tau^{i}_{\sigma\sigma'}d_{\sigma'} \ , \\
    \vec{S}_{k,k'} & = \frac{1}{2}\sum_{i{=}x,y,z}\sum_{kk'\sigma\sigma'}c_{k\sigma}^\dag\tau^{i}_{\sigma\sigma'}c_{ k'\sigma'} \ ,
    \end{align}
    \label{eq: supmat spinor notation}
\end{subequations}
are the spin operators of the dot and lead averaged over all possible scattering channels, whereas the exchange scattering coefficients are
\begin{equation}
\label{eq: supmat SW exchange scattering}
J= \dfrac{2U\left|t_{\text{D}}\right|^2}{(\epsd + U)(-\epsd)} \ , \ \ 
W_{\alpha\alpha'} = 
\dfrac{(2\epsd + U)\left|t_{\text{D}}\right|^2}{2(\epsd + U)(-\epsd)} \ .
\end{equation}
For $\epsilon_{_\text{D}}{=}{-}U/2$, $J{>}0$ and we can identify the dot-lead Kondo singlet formation because of the antiferromagnetic interaction. This leads to a state of the form
\begin{equation}
    \ket{\psi} = \frac{1}{\sqrt{2}}\left(\ket{\uparrow}_\text{D}\otimes\ket{\downarrow}_\text{lead} - \ket{\downarrow}_\text{D}\otimes\ket{\uparrow}_\text{lead}\right) \ ,
\end{equation} 
where $\ket{\sigma}_\text{lead}$ describes the collective lead's state with total magnetic moment $\sigma$. The dot's reduced density matrix of this state reads
\begin{equation}
    \rho_\text{D} = \frac{1}{2}\left(\ketbra{\uparrow}{\uparrow}_\text{D}+\ketbra{\downarrow}{\downarrow}_\text{D}\right) = \frac{1}{2}\sum_\sigma\ketbra{\sigma}{\sigma}_\text{D} \ .
\end{equation}
The first effect discussed in the main text for system (1) corresponds to the left dot forming such a Kondo singlet with its own lead, while the right dot is empty [cf. Fig.~\ref{fig: figure 2}(a,i) and (b)]. Therefore, the density matrix of the double-dot system reads
\begin{align}
    \rho_\text{DD} & = \rho_\text{L}\otimes\rho_\text{R} =  \frac{1}{2}\sum_\sigma\ketbra{\sigma}{\sigma}_\text{L}\otimes\ketbra{0}{0}_\text{R} \nonumber  \\
     & = \frac{1}{2}\sum_\sigma\ketbra{\sigma,0}{\sigma,0} \ .
\end{align}
We read the density matrix elements, used as order parameters, obtaining the values as reported in column (i) of Table~\ref{table 1} of the main text.

As a last note regarding the Schrieffer-Wolff transformation, we would like to mention that by plugging in the SWT Hamiltonian~\eqref{eq:KondoSingle} into a standard calculation for conductance, it is additionally possible to estimate the gap opening due to the Kondo singlet formation
\begin{equation}
\label{eq: supmat Kondo temperature quantum dot with one lead}
\Delta_\text{K}= \sqrt{\frac{\pi|t_\text{D}|^2Ud_0}{2}}\exp\Biggl(\dfrac{\epsd(\epsd+U)}{2|t_\text{D}|^2Ud_0}\Biggr) \ ,
\end{equation}
where $d_0$ is the lead's density of states, see Ref.~\cite{bruus_many-body_2004} for a detailed derivation. We use this gap in the main text to compare it with the gap that RKKY terms engender. %In the main text, we estimate the coupling terms $J$ and Kondo gaps $\Delta_\text{K}$ for both standard Kondo and Kondo cat effects analogously.

%%%%%%%%%%%%%%%%%% Exact diagonaliozation and SWT

\section{Exact diagonalization and SWT of the closed cavity-double-dot system; case (2)}
In the main text, we discuss the charge stability diagram [cf.~Fig.~\ref{fig:model}(d) in the main text] of the closed-system Hamiltonian 
\begin{equation}
H_{\text{CDD}}{=}\sum_{\ell}\left(H_{\ell}{+}H_{\text{tun}}^{C{-}\ell}\right){+}H_{\text{C}}^{\delta} \ ,
\label{eq: supmat equation closed cavity-double-dot system}
\end{equation}
with the dots' Hamiltonian $H_{\ell}{=}\sum_\sigma^{\phantom \dag}\epsilon_{\ell}^{\phantom\dag}n_{\ell\sigma}^{\phantom\dag}{+}U_\ell^{\phantom\dag} n_{\ell\uparrow}^{\phantom\dag}n_{\ell\downarrow}^{\phantom\dag}$, the cavity-dots tunneling terms $H_{\text{tun}}^{C{-}\ell}{=}\sum_{\ell  j\sigma}\tilde{t}_{\ell}\ddag_{\ell\sigma} \cpdag_{j\sigma}{+}\text{H.c.}$, and the cavity Hamiltonian $H_{\text{tun}}^{C{-}\ell}{=}\sum_{\ell  \sigma}\tilde{t}_{\ell}\ddag_{\ell\sigma} \cpdag_{\sigma}{+}\text{H.c.}$. We commonly truncate the latter's Hilbert space to a single energy level and consider the detuned cavity regime $n_\text{C}{\approx}0$.

Here, we apply exact diagonalization and SWT to the configurations (iii)-(v), cf.~Fig.~\ref{fig: figure 2}(a) in the main text. These methods yield a perturbative (in the lead-impurity coupling) analytical treatment of the system, which we  compare with the numerical (exact) analysis in the main text.
\paragraph{Cavity-mediated RKKY.}
For $\epsilon_{_\text{L}}{\approx}\epsilon_{_{\text{R}}}{\approx}{-}U/2$, each dot is approximately singly occupied. According to the SWT [cf.~Sec.~\ref{sec:singleDot}], each dot can separately form a singlet with its respective lead, and the open cavity-double-dot effective model~~\eqref{eq: supmat equation closed cavity-double-dot system} reads
\begin{equation}
        H \approx H_{\text{CDD}} + H_{\text{leads}} + J\sum_{\ell kk'}\vec{S}_{\ell}\cdot\vec{S}_{\ell k,\ell k'} \ ,
        \label{eq: detuned cavity resulting model}
    \end{equation}
    where the spinor operators are defined in Eq.~\eqref{eq: supmat spinor notation}. Thus, each dot's density matrix is a mixture of the form
\begin{equation}
\rho_\ell = \frac{1}{2}\sum_\sigma\ketbra{\sigma}{\sigma} \ ,
\end{equation}
and the double-dot system forms the mixture
\begin{equation}
    \rho_\text{DD} = \rho_\text{L}\otimes\rho_\text{R} \approx \frac{1}{4}\sum_{\sigma,\sigma'}\ketbra{\sigma,\sigma'}{\sigma,\sigma'} \ ,
\end{equation}
see Fig.~\ref{fig: figure 2}(a,iii) of the main text. We read the density matrix elements, used as order parameters, obtaining the values as reported in column (iii) of Table~\ref{table 1} of the main text. As two separated Kondo effects form, a gap opens which is twice the size of the single-dot Kondo gap~\cite{bruus_many-body_2004}, [cf.~Eq.~\eqref{eq: supmat Kondo temperature quantum dot with one lead}]
\begin{equation}
\label{eq: supmat Kondo temperature quantum dot with one lead}
\Delta_\text{K}= \sqrt{2\pi|t_\text{D}|^2Ud_0}\exp\Biggl(\dfrac{\epsd(\epsd+U)}{2|t_\text{D}|^2Ud_0}\Biggr) \ .
\end{equation}
As a next step, we include the presence of the cavity in our description. With exact diagonalization, we determine the ground state of the closed system $H_{\text{CDD}}$ in the detuned-cavity and $\epsilon_{_\text{L}}{\approx}\epsilon_{_\text{R}}{\approx}{-}U/2$ region
\begin{align}
 \ket{0}_\text{C}\otimes\frac{1}{\sqrt{2}}\left(\ket{\uparrow,\downarrow}-\ket{\downarrow,\uparrow}\right) \ ,
  \label{eq: supmat dot singlet state}
        \end{align}
        where with $\ket{s}_\text{C}$ we denote the configuration of the cavity. This state corresponds to a cavity-mediated singlet that forms on the two dots and the dot-dot reduced density matrix reads 
\begin{equation}
\rho_\text{DD}\approx\frac{1}{2}\sum_{\sigma}\left(\ketbra{\sigma,\bar{\sigma}}{\sigma,\bar{\sigma}}-\ketbra{\sigma,\bar{\sigma}}{\bar{\sigma},\sigma}\right)\, ,
    \label{eq: RKKY singlet}
\end{equation}
see Fig.~\ref{fig: figure 2}(a,iv). We read the density matrix elements obtaining the values as reported in column (iv) of Table~\ref{table 1} of the main text. As this configuration is nondegenerate, the dots decouple from the leads, hence, the term ``exchange blockade''. 
    The first excited states, in this configuration, form a triplet 
       \begin{align}
 & \ket{0}_\text{C}\otimes\frac{1}{\sqrt{2}}\left(\ket{\uparrow,\downarrow}+\ket{\downarrow,\uparrow} \right) \ , \nonumber \\
  & \ket{0}_\text{C}\otimes\frac{1}{\sqrt{2}}\ket{\uparrow,\uparrow}\ ,  \\
   & \ket{0}_\text{C}\otimes\frac{1}{\sqrt{2}}\ket{\downarrow,\downarrow}\ . \nonumber
  \label{eq: supmat dot triplet state}
        \end{align}
With a degenerate perturbation theory procedure, we find that this triplet is split from the ground state with energy 
\begin{equation}
    \Delta_\text{RKKY} \approx 48\tilde{t}_{_\text{D}}^4{/}\left(U{+}2\epsilon_{_\text{C}}\right)^3 \ .
    \end{equation}
    As a result, we can trace out the cavity, as it always dwells in the $\ket{0}_\text{C}$ state, to obtain an effective description of the closed system Hamiltonian
    \begin{equation}
        H_{\text{CDD}}\to H_{\text{CDD}}^{\text{AF}}\approx\Delta_\text{RKKY}\vec{S}_\text{L}{\cdot}\vec{S}_\text{R} \, ,
    \end{equation}
    with antiferromagnetic interaction $\Delta_\text{RKKY}$.  
\paragraph{Kondo cat.}
As shown in Fig.~\ref{fig:model}(d), the coupling to the cavity opens a gap in the $\epsilon_{_\text{L}}{\approx}\epsilon_{_\text{R}}{\gtrsim}0$ region and the total occupation of the double dot system is $n_\text{DD}{\approx}1$ in the exchange blockade regime. Using exact diagonalization, we determine the spin-degenerate ground state of $H_{\text{CDD}}$ to be
\begin{align}
 \ket{0}_\text{C}\otimes\frac{1}{\sqrt{2}}\left(\ket{\sigma,0}+\ket{0,\sigma}\right)  \ .
  \label{eq: supmat Kondo cat ground}
        \end{align}
 Using degenerate perturbation theory, we calculate the energy of this ground state (second-order correction term) $\epsilon_{_\text{D}} - \Delta_\text{EXC}$, where,as in the main text, $\Delta_\text{EXC}{=}2\tilde{t}_{_\text{D}}^2/(\epsilon_{_\text{C}}-\epsilon_{_\text{D}})$.
 The first excited states are the empty configurations
 \begin{equation}
 \ket{0}_\text{C}\otimes\frac{1}{\sqrt{2}}\left(\ket{\sigma,0}+\ket{0,\sigma}\right) \ , 
 \end{equation}
 with zero energy, and doubly-occupied as in Eq.~\eqref{eq: supmat dot singlet state}, with energy $2\epsilon_{_\text{D}}-\Delta_{\text{RKKY}}$. As discussed in the main text, the exchange blockade region, therefore, forms for
 \begin{align}
\epsilon_{_\text{D}}- \Delta_\text{EXC} & < 0 \nonumber \\
\epsilon_{_\text{D}}-\Delta_\text{EXC} & <  \min(2\epsilon_{_\text{D}} - \Delta_\text{K}, 2\epsilon_{_\text{D}} - \Delta_\text{RKKY}) \ ,
 \end{align}
 where in the second relation we take into account the competition between standard Kondo effects and RKKY hybridization in the double occupation regime $n_\text{DD}{\approx}2$. The boundaries of these two relations are delineated in Fig.~\ref{fig: figure 2}(d) using dashed grey lines. From Eq.~\eqref{eq: supmat Kondo cat ground}, we conclude that in the $n_\text{D}{\approx}1$ the dot-dot configuration is spin-degenerate and with a nonvanishing magnetic moment. Therefore, the degenerate ``ground state'' can reduce its energy by forming a singlet of Kondo nature. The resulting dot-dot density matrix is the mixture
\begin{equation}
    \rho_{\text{DD}} \approx \frac{1}{4}\sum_\sigma\left(\ket{\sigma,0}+\ket{0,\sigma}\right)\left(\bra{\sigma,0}+\bra{0,\sigma}\right) \ ,
\end{equation} see Figs.~\ref{fig: figure 2}(a,v). We read the density matrix elements, used as order parameters, obtaining the values as reported in column (v) of Table~\ref{table 1} of the main text. This singlet, which we call a Kondo cat. The Kondo gap further reduces the system's energy  in the $n_{\text{DD}}{\approx}1$. Therefore, we expect Kondo cat formation in the gap independently of the cavity-dots coupling parameter $\tilde{t}_\text{D}$. The exchange scattering coefficients and Kondo gap [Eq.~\eqref{eq: supmat SW exchange scattering} and Eq.~\eqref{eq: supmat Kondo temperature quantum dot with one lead}, respectively, for the single impurity Anderson model] can be determined analogously to the procedure followed in Ref.~\cite{bruus_many-body_2004}.
%%%%%%%%%%%%%%%%% VMPS algorith for degenerate states
\section{VMPS algorithm for degenerate states}
For a system with $n$ degenerate ground states $\left\{\ket{\psi_i}\right\}_{i=1}^n$, the VMPS algorithm randomly finds any  superposition of the ground states $\sum_{i{=}1}^n\alpha_i\ket{\psi_i}$, with $\alpha_i$ unknown, which depends on numerical details and initial conditions. As mentioned in the main text, the full double dot system (1) has three degenerate states in the regime $\epsilon_{_\text{L}}{\approx}\epsilon_{_\text{R}}{\approx}{-}U/2$, namely, the three triplet states $\ket{\psi_T^0}, \ket{\psi_T^+}, \ket{\psi_T^-}$~\cite{ong_generalized_2011}. Therefore, the VMPS algorithm finds a (normalized) state of the form
\begin{equation}
    \ket{\psi} = \alpha_T^0\ket{\psi_T^0} + \alpha_T^+\ket{\psi_T^+} + \alpha_T^-\ket{\psi_T^-} \ .
    \label{eq: ground state degenerate VMPS}
\end{equation}
In the states $\ket{\psi_T^0}, \ket{\psi_T^+}, \ket{\psi_T^-}$, the two dots form either an antiferromagnetic triplet
\begin{equation}
\rho_\text{DD}\approx\frac{1}{2}\sum_{\sigma}\left(\ketbra{\sigma,\bar{\sigma}}{\sigma,\bar{\sigma}}+\ketbra{\sigma,\bar{\sigma}}{\bar{\sigma},\sigma}\right)\ ,
    \label{eq: RKKY singlet supmat}
\end{equation} 
or are in a ferromagnetic configuration
\begin{equation}
\rho_\text{DD}\approx\ketbra{\sigma,\sigma}{\sigma,\sigma} \ .
    \label{eq: ferromagnetic supmat}
\end{equation}
Consequently, the matrix elements $\bra{\sigma,\sigma'}\rho_{\text{DD}}\ket{\sigma,\sigma'}$ or $\bra{\sigma,\sigma}\rho_{\text{DD}}\ket{\sigma,\sigma}$ obtained from the VMPS ground state include the information related to the numerical (random) coefficients $\alpha_T^0,\alpha_T^+,\alpha_T^-$. By plotting the sum of the antiferromagnetic and ferromagnetic contributions
\begin{equation}
    \bra{\sigma,\sigma'}\rho_{\text{DD}}\ket{\sigma,\sigma'} + \bra{\uparrow,\uparrow}\rho_{\text{DD}}\ket{\uparrow,\uparrow} + \bra{\downarrow,\downarrow}\rho_{\text{DD}}\ket{\downarrow,\downarrow} \ ,
\end{equation}
we eliminate this numerical parameter due to the imposed normalization constraint
\begin{equation}
    \left|\alpha_T^0\right|^2 + \left|\alpha_T^+\right|^2 + \left|\alpha_T^-\right|^2 = 1 \ .
\end{equation}
\end{document}